\documentclass[reprint,nofootinbib, amsmath,amssymb,aps,floatfix]{revtex4-2}

\usepackage{graphicx}% Include figure files
\usepackage{dcolumn}% Align table columns on decimal point
\usepackage{bm}% bold math
\usepackage[colorlinks=true,pdfstartview=Fit]{hyperref}

\usepackage{physics}		% for derivatives, differentials, etc.
\DeclareMathOperator {\e}{e}	% e for exponential should be roman
\usepackage{bm}			% bold math

\usepackage[dvipsnames]{xcolor}   
\usepackage[normalem]{ulem}		
\begin{document}

\title{The metastable Mpemba effect corresponds to a non-monotonic temperature dependence of extractable work}

\author{Rapha\"el Ch\'etrite}
\affiliation{Laboratoire J A Dieudonné, UMR CNRS 7351, Université de Nice Sophia
Antipolis, Nice, France}
\email{raphael.chetrite@unice.fr}

\author{Avinash Kumar, John Bechhoefer}
\affiliation{Dept. of Physics, Simon Fraser University, Burnaby, British Columbia, V5A 1S6, Canada}

\begin{abstract}
The Mpemba effect refers to systems whose thermal relaxation time is a non-monotonic function of the initial temperature.  Thus, a system that is initially hot cools to a bath temperature more quickly than the same system, initially warm.  In the special case where the system dynamics can be described by a double-well potential with metastable and stable states, dynamics occurs in two stages:  a fast relaxation to local equilibrium followed by a slow equilibration of populations in each coarse-grained state.  We have recently observed the Mpemba effect experimentally in such a setting, for a colloidal particle immersed in water.  Here, we show that this \textit{metastable Mpemba effect} arises from a non-monotonic temperature dependence of the maximum amount of work that can be extracted from the local-equilibrium state at the end of Stage 1.
\end{abstract}

\maketitle

\section{Introduction}
\label{sec:intro}

A generic consequence of the second law of thermodynamics is that a system, once perturbed, will tend to relax back to thermal equilibrium.  Such relaxation is typically exponential.  To understand why, consider energy relaxation and recall that the heat equation,
\begin{align}
	\pdv{T}{t} = \kappa \nabla^2 T(\bm{r},t) \,,
\label{eq:heat-diffusion}
\end{align}
for the temperature field $T$ at position $\bm{r}$ and time $t$ and thermal diffusivity $\kappa$ has solutions that can be written in the form\footnote{
We begin the eigenfunction expansion at $m=2$ to be consistent with the analogous expansion of the Fokker-Planck solution in \citet*{lu17}.}
\begin{align}
	T(\bm{r},t) = T_\infty(\bm{r}) + \sum_{m=2}^\infty a_m \, v_m(\bm{r}) \e^{-\lambda_m t} \,.
\label{eq:heat-diffusion-inf-series}
\end{align}
Here, $T_\infty(\bm{r})$ is the static temperature-field solution of Equation~\eqref{eq:heat-diffusion}; it  must account for boundary conditions.  For $t \to \infty$, an arbitrary initial condition $T(\bm{r},0)$ will relax to this state.  In Equation~\eqref{eq:heat-diffusion-inf-series}, the $v_m(\bm{r})$ are spatial eigenfunctions, with corresponding eigenvalues $\lambda_m$ and coefficients $a_m$, which represent the projections of the field $[T(\bm{r},0)-T_\infty(\bm{r})]$ onto the corresponding eigenfunction.  For long but finite times, all but the slowest eigenmode will have decayed, and the temperature is, approximately,
\begin{align}
	T(\bm{r},t) \approx T_\infty (\bm{r}) + a_2 \, v_2(\bm{r}) \e^{-\lambda_2 t} \,,
\label{eq:heat-diffusion-long-time}
\end{align}
which, indeed, shows a simple exponential decay to $T_\infty (\bm{r})$ for a probe at a fixed position $\bm{r}$.

Although exponential decays are typical, anomalous, non-exponential relaxation is also  encountered.  Large objects, for example, may have an asymptotic time scale $\lambda_2^{-1}$ that exceeds experimental times, so that it is not possible to wait ``long enough."  Similarly, glassy systems and other complex materials may have a spectrum of exponents for mechanical and dielectric relaxation that have not only very long time scales but also many closely spaced values that are not resolved as a sequence of exponentials.  Rather, they can collectively combine to approximate a power-law or even logarithmic time decay, with specific details that depend on the history of preparation \cite{amir12}.

Another class of anomalous systems shows unexpectedly \textit{fast} relaxation in certain circumstances.  The best-known of these is the observation that, occasionally, a sample of hot water may cool and begin to freeze more quickly than a sample of cool or warm water prepared under identical conditions.  Based on the scenario of exponential relaxation sketched above, one's naive intuition is that a hotter system will have to ``pass through'' all intermediate temperatures and thus take longer to equilibrate.  More succinctly, the observation is that, in some systems, the equilibration time is a non-monotonic function of the initial temperature:  the time for a system initially in equilibrium at a given temperature takes to cool and reach equilibrium with the bath temperature does not always increase with initial temperature.  

While observations of this phenomenon date back two millennia to the ancient Greeks \cite{webster1923,ross81}, its modern study began with observations by Mpemba in the 1960s \cite{mpemba69}.  The effect has since been observed in systems such as manganites \cite{chaddah10}, clathrates \cite{ahn16}, polymers \cite{lorenzo06,hu18} and predicted in simulations of other systems, including carbon nanotube resonators \cite{greaney11}, granular fluids \cite{lasanta17}, and spin glasses \cite{baity19}.  In all these \textit{Mpemba effects}, the relaxation time  shows a surprisingly complicated dependence on the deviation of initial temperature from equilibrium:  increasing and then decreasing, and in some cases increasing again with increasing deviation.  The relaxation time thus does not increase monotonically with the deviation from equilibrium, as one might naively expect.

One challenge in studying Mpemba effects is that the systems where they have been observed or predicted have been rather complicated, with many possible explanations for the effect.  The explanations tend to be complicated and specific to a particular system.  Even water is not as simple as it might seem:  proposed mechanisms include evaporation \cite{kell69,vynnycky10,mirabedin17}, convection \cite{vynnycky15}, supercooling \cite{auerbach95}, dissolved gases \cite{wojciechowski88}, and effects arising from hydrogen bonds \cite{zhang14}.

In an effort to understand the Mpemba effect more generically, Lu and Raz recently proposed an explanation that is linked to the structure of eigenfunction expansions such as that in Equation~\eqref{eq:heat-diffusion-inf-series} \cite{lu17}.  Their work was formulated for mesoscopic systems that are in the classical regime yet are small enough that thermal fluctuations make an important contribution to their dynamics.  Such systems may be described by master equations and Fokker-Planck equations, for finite and continuous state spaces, respectively \cite{risken89,van-kampen07,hanggi82,gardiner09,seifert12}.  For the latter, the Fokker-Planck equation describes the evolution of the probability density function $p(\bm{x},t)$ for a system described by a state vector $\bm{x}(t)$.\footnote{
In a many-body system, the dimension of $\bm{x}$ can be very large.}
Its structure is similar to that of Equation~\eqref{eq:heat-diffusion}: its linearity implies that solutions are also described by an infinite-series, eigenfunction expansion similar to that in Equation~\eqref{eq:heat-diffusion-inf-series}.  The essence of Lu and Raz's explanation is that the projection of the initial state $p(\bm{x},0)$ -- a Gibbs-Boltzmann distribution corresponding to an initial temperature $T$ -- onto the slowest eigenfunction, $a_2$ can be non-monotonic in $T$,  or, equivalently, in $\beta^{-1} \equiv k_\textrm{B}T$, where $k_\textrm{B} \equiv 1$ (in our units) is Boltzmann's constant.   Such a consequence implies a Mpemba effect because the long-time limit for the probability density function has the same form as Equation~\eqref{eq:heat-diffusion-long-time}:
\begin{align}
	p(\bm{x},t) \approx g_{\beta_\textrm{b}}(\bm{x}) + a_2(\beta,\beta_\textrm{b}) \, 
		v_{2,\beta_\textrm{b}}(\bm{x}) \e^{-\lambda_2 t} \,,
\label{eq:pdf-long-time}
\end{align}
with $g_{\beta_\textrm{b}}(\bm{x})$ the Gibbs-Boltzmann distribution for the system at a temperature $T_\textrm{b}$ corresponding to the surrounding thermal bath with which the system is in contact and can exchange energy.  The coefficient $a_2$ is a function of \textit{both} the initial temperature and bath temperature:
\begin{align}
	a_2(\beta,\beta_\textrm{b}) = \int \dd{\bm{x}} g_\beta (\bm{x}) u_{2,\beta_\textrm{b}}(\bm{x})  \,,
\label{eq:pdf-a2}
\end{align}
where the initial state $p(\bm{x},0)$ is assumed to be in equilibrium at a higher temperature $\beta^{-1}$ and where $u_{2,\beta_\textrm{b}}(\bm{x})$ is the \textit{left eigenfunction} of the Fokker-Planck operator, which is the \textit{dual-basis element} corresponding to the \textit{right eigenfunction} $v_{2,\beta_\textrm{b}}(\bm{x})$ of the same Fokker-Planck operator.  Both $u_2$ and $v_2$ are calculated for the Markovian Langevin dynamics associated with white noise whose covariance is set by the bath temperature, $\beta_\textrm{b}^{-1}$.  We need to distinguish between left and right eigenfunctions because the operator generating Fokker-Planck dynamics is not self-adjoint, in contrast to the operator generating the heat-diffusion dynamics discussed in Equation~\eqref{eq:heat-diffusion}.  The Mpemba effect then translates to the non-monotonicity of $a_2$ as a function of the initial temperature $\beta^{-1}$:  If a high-temperature initial condition has a smaller coefficient $a_2$, then, in the long-time limit, the system will be closer to equilibrium than a cool-temperature initial condition with larger $a_2$.  This non-monotonicity in $a_2$ is easier to establish than the non-monotonicity of equilibration times that defines the Mpemba effect.  The latter requires either an experiment or, at the very least, repeated numerical solution of the full Fokker-Planck equation.

Inspired by the scenario proposed by \citet{lu17}, we have explored the Mpemba effect in a simple, mesoscopic setting that -- unlike previous work -- lends itself to quantitative experiments that straightforwardly connect with theory \cite{kumar20}.  In particular, we explored the motion of a single micron-scale colloidal particle immersed in water and moving in a tilted double-well potential.  The one-dimensional (1D) state space consists of the position $x(t)$ of the particle.  By choosing carefully the tilt of the potential, along with the energy-barrier height and the offset (asymmetry) of the double-well potential within a box that confines the particle motion at high temperatures, we could demonstrate convincingly the existence of the Mpemba effect and measure the non-monotonic temperature dependence of the $a_2$ coefficient.  We even found conditions where $a_2=0$.  At such a point, the slowest relaxation dynamics is $\sim \e^{-\lambda_3 t}$, implying an \textit{exponential speed-up} over the generic relaxation dynamics, $\sim \e^{-\lambda_2 t}$.  This \textit{strong Mpemba effect} had been predicted by \citet{klich19}.

Although our recent experimental work gives strong support to the basic scenario proposed by Lu and Raz, it does not offer good physical insight into the conditions needed to produce or observe the Mpemba effect.  
% What kinds of potentials can lead to the effect?  
What physical picture corresponds to the anomalous temperature dependence of the $a_2$ coefficient?  In this Brief Research Report, we offer a more physical interpretation of the Mpemba effect explored in our previous work.

\section{Thermalization in a double-well potential with metastability }

\begin{figure}[h!]
\begin{center}
	\includegraphics[width=8cm]{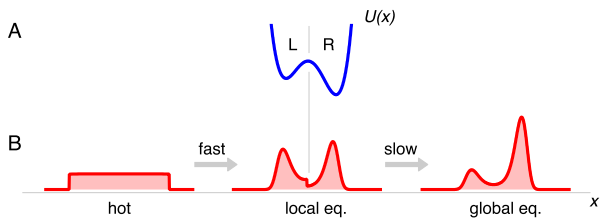}
\end{center}
\caption{Two-stage dynamics  (A) Tilted double-well potential $U(x)$ with coarse-grained states $\{L,R\}$.  The potential includes a box (not shown).  (B)  Evolution of the probability density function for position: a high-temperature equilibrium initial state $g_\beta (x)$ (left) has a fast relaxation to a local equilibrium state $\rho_{\beta,\beta_\textrm{b}}^\textrm{leq}(x)$ (middle) and a slow relaxation to global equilibrium $g_{\beta_\textrm{b}}(x)$ at the colder bath temperature (right).}
\label{fig:localEq}
\end{figure}

\vspace{1em}
A common feature of experiments showing Mpemba effects is that they involve a temperature \textit{quench}: the system is cooled very rapidly.  We model this situation by making the high-temperature initial state an initial condition for dynamics that take place entirely in contact with a bath of fixed temperature.  In effect, the quench is infinitely fast.  The thermalization dynamics are then given by the Langevin equation
\begin{align}
	\dot{x} = -\gamma U'(x)+\sqrt{ 2\gamma \beta_\textrm{b}^{-1}} \, \eta,
\end{align}
with $\gamma$ a friction coefficient and $\eta(t)$ Gaussian white noise modeling thermal fluctuations from the bath, with $\langle \eta(t) \rangle = 0$ and $\langle \eta(t) \, \eta(t') \rangle = \delta(t-t')$.  The noise-strength $2\gamma \beta_\textrm{b}^{-1}$ enforces the fluctuation-dissipation relation \cite{van-kampen07,gardiner09}. The potential $U(x)$ is a double-well potential with barrier height $E_0 \gg {\beta_\textrm{b}}^{-1}$ and two coarse-grained states, denoted $L$ and $R$ in Figure~\ref{fig:localEq}A.  The range of particle motions is also constrained to a finite range; the potential is implicitly infinite at the extremities.  By tilting the potential, one state has a higher energy than the other (difference is $\Delta E$) and becomes a toy model for the water-ice phase transition.  However, the energy barrier $E_0$, while high enough that the two states are well defined, is also low enough that many transitions over the barrier are observed during a typical experiment.  

Figure~\ref{fig:localEq} illustrates the case studied by \citet{kumar20}, with (a) showing the potential and (b) the dynamics of a quench from a high temperature.  With a moderately high barrier, both wells have significant probability for the equilibrium state $g_{\beta_\textrm{b}}(x)$ (Figure~\ref{fig:localEq}B, right).  For $U(x)$, the barrier $E_0 = 2.0$, the energy difference between states is $\Delta E = 1.3$, and the hot temperature $\beta_\textrm{h}^{-1} = 1000$; all quantities are multiplied by $\beta_\textrm{b}$ and are, hence, dimensionless.  The separation between wells was 80 nm.

At a temperature corresponding to $\beta^{-1}$, the equilibrium free energy of the system is 
\begin{align}
	F_{\beta}^\textrm{eq} \equiv -\beta^{-1} \ln \left[
	\int_{-\infty}^{+\infty} \dd{x} \exp \left( -\beta U(x) \right) \right] \,.
\label{eq:nrllibre}
\end{align}
and the corresponding equilibrium Gibbs density is
\begin{align}
	g_{\beta}(x) \equiv \exp \left[ -\beta \left( U(x)-F_{\beta}^\textrm{eq} \right) \right] \,,
\label{eq:gibbs}
\end{align}

The metastability of $U$ means that the system evolves on two very different time scales:  

\textit{Stage 1} is a fast relaxation to local equilibration. The initial, high-temperature Gibbs density rapidly evolves to a state that is at local equilibrium with respect to the bath temperature.  A local equilibrium is a density that is similar locally to $g_{\beta_\textrm{b},}$ but with altered fractions of systems in the left or right wells.  Using Bayes' theorem, we can write such a local-equilibrium state as 
\begin{widetext}
\begin{align*}
	\rho_{\beta,\beta_\textrm{b}}^\textrm{leq}(x) 
	& =\mathbb{P}\left(\textrm{ be in the left well at } \beta \right)
	\mathbb{P}\left(\left.x\right|\textrm{ be in the left well at } \beta_\textrm{b} \right) \\
	& \quad +\mathbb{P}\left(\textrm{ be in the right well at } \beta \right)
	\mathbb{P}\left(\left.x\right|\textrm{ be in the right well at } \beta_\textrm{b} \right) \,.
\end{align*}
More precisely, the local $\beta,\beta_\textrm{b}$ equilibrium  is the density 
\begin{align}
	\rho_{\beta,\beta_\textrm{b}}^\textrm{leq}(x) 
	&= \begin{cases}
		a_\textrm{L}  \left( \frac{g_{\beta_\textrm{b}}(x)}
		{\int_{-\infty}^0 \dd{x'} g_{\beta_\textrm{b}}(x')} \right) 
	& \qquad x<0 	\quad	\textrm{(left well)} \,, \\[12pt]
	a_\textrm{R}  \left( \frac{g_{\beta_\textrm{b}}(x)}
		{\int_{0}^{\infty} \dd{x'} g_{\beta_\textrm{b}}(x')} \right) 
	& \qquad x>0 	\quad 	\textrm{(right well)} \,,
	\end{cases}
\label{eq:leq}
\end{align}
with $0 \le a_\textrm{L} \le 1$.  Choosing $a_\textrm{L} + a_\textrm{R} =1$ ensures normalization of the probability density.
\end{widetext}

In a fast quench, we assume that the fraction of initial systems at equilibrium at the higher temperature $\beta^{-1}$ is unchanged when local equilibrium is established.  Essentially, we ignore the diffusion of trajectories that start on one side of the barrier and end up on the other at the end of the transient.  In this approximation, the fraction that ends up in each well corresponds to that of the initial state, $g_{\beta}$.  Thus,
\begin{align}
	a_\textrm{L} = \int_{-\infty}^0 \dd{x'} g_{\beta}(x') \qquad \text{and} \qquad
	a_\textrm{R} = \int_0^{\infty} \dd{x'} g_{\beta}(x') \,.
\label{eq:leq1}
\end{align}   
As shown in Figure~\ref{fig:localEq}B, center, the local-equilibrium distribution $\rho_{\beta,\beta_\textrm{b}}^\textrm{leq}(x)$ is discontinuous at $x=0$; higher barriers will reduce the discontinuity, of order $\e^{-\beta_\textrm{b} E_0} \ll 1$. 

\textit{Stage 2} is a final relaxation to global equilibrium on a slow time scale:  the overall populations in each well (coarse-grained state) change, and the density converge to the Gibbs density $g_{\beta_\textrm{b}}$.  Local equilibrium is maintained during the evolution, which is illustrated schematically in Fig~\ref{fig:localEq}b.  In this \textit{metastable regime}, the equilibration time was analyzed by Kramers long ago \cite{kramers40,van-kampen07,hanggi90,berglund13}.  It also corresponds to the limit of Equation~\eqref{eq:pdf-long-time}; as a result, the final relaxation is exponential, with decay rate $\lambda_2$.

\section{Metastable Mpemba effect}
\label{sec:metastable}

\vspace{1em}

Given this scenario of thermal relaxation as a two-stage process, we can readily understand how the Mpemba effect can occur.  The idea is to follow the dynamics in the function space of all admissible probability density functions $p(x,t)$.  If we expand the solution in eigenfunctions analogously to Equation~\eqref{eq:heat-diffusion-inf-series}, we see that the infinite-dimensional function space is spanned by the eigenfunctions.  To visualize the motion, we project it onto the 2D subspace spanned by the eigenfunctions $v_2(x)$ and $v_3(x)$.  The system state is then characterized as a parametric plot of the amplitudes $a_2(t)$ and $a_3(t)$.   Animations from a 3D projection spanning $a_2$--$a_3$--$a_4$ are available in the supplementary material.  A similar geometric plot was used to explore quenching in an anti-ferromagnetic Ising spin system in \cite{klich19}.

\begin{figure}[h!]
\begin{center}
	\includegraphics[width=8cm]{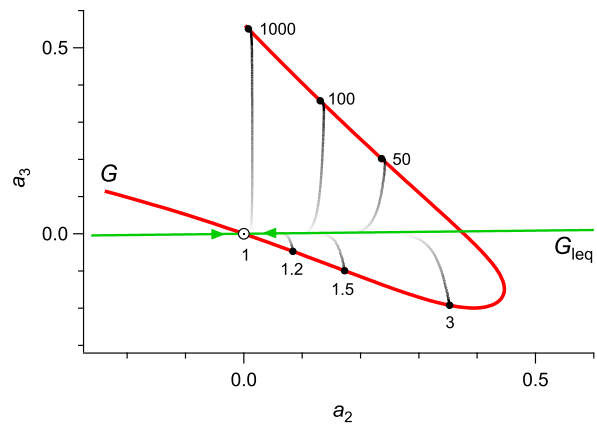}
\end{center}
\caption{Probability-density dynamics in the plane defined by the $a_2$ and $a_3$ coefficients.  The red curve $G$ denotes the set of equilibrium densities, the green curve $G_\textrm{leq}$ the set of local-equilibrium densities.  Arrows indicate the slow relaxation along $G_\textrm{leq}$ to global equilibrium, at the intersection with $G$ (denoted by the large hollow marker with a dot at its center at $T=1$).  Gray lines denote the rapid relaxation from an initial condition (temperature relative to the bath indicated by a marker along $G$).  The time progression of $p(x,t)$, projected onto the $a_2$--$a_3$ plane, is from dark to light.  Curves are calculated from the double-well potential described in Kumar and Bechhoefer, with domain asymmetry $\alpha=3$ (see \cite{kumar20} for definitions).}
\label{fig:alpha3plot}
\end{figure}

Figure~\ref{fig:alpha3plot} shows the geometry of trajectories.  They are organized about two static, 1D curves, labeled $G$ and $G_\textrm{leq}$.  The red curve ($G$) represents the set of all equilibrium Gibbs-Boltzmann densities, $g_\beta$, for $0 \le \beta < \infty$.  It is sometimes known as the \textit{quasi-static locus}.  The green curve ($G_\textrm{leq}$) represents the set of all local-equilibrium densities of the form of Equation~\eqref{eq:leq}, as parametrized by $a_\textrm{L} \in [0,1]$.  Both curves are represented as 2D parametric plots but lie in the full infinite-dimensional space.  Both $G$ and $G_\textrm{leq}$ have finite length, in general.  (The entire length is not shown in the figure.)  The two curves intersect at $a_2=a_3=0$, which describes the global equilibrium $g_{\beta_\textrm{b}}$ with respect to the bath (large hollow marker with dot).  The apparent crossing near $a_2 \approx 0.4$ is spurious, as the 3D projections in the supplement show.

The dynamical trajectories are represented by the variously shaded gray curves.  At time $t=0$, the systems are in equilibrium along the red curve at a variety of temperatures $\{1, 1.2, 1.5, 3, 50, 100, 1000\} \times T_\textrm{b}$, which are indicated by black markers.  The curves then move rapidly towards the green curve (local equilibrium).  The time course is suggested by the dark-to-light gradient.  Once they reach the vicinity of $G_\textrm{leq}$, they closely follow this green curve back to the global-equilibrium state.

Within this representation, we note the ``arrival point'' of each trajectory when it ``hits'' $G_\textrm{leq}$.  For small temperatures (1, 1.2, 1.5, 3), the distance between this arrival point and the global-equilibrium state increases monotonically with $\beta$.  For larger temperatures (50, 100, 1000), however, the distance decreases until, at $T=1000 T_\textrm{b}$, it nearly vanishes (denoting the strong Mpemba effect).  Along $G_\textrm{leq}$, the system is in the limit described by Equation~\eqref{eq:pdf-long-time} and relaxes exponentially to global equilibrium.  Relaxation along $G_\textrm{leq}$ therefore \textit{must} be monotonic with the distance away from global equilibrium.  Trajectories that arrive along this curve that are farther from global equilibrium will take longer to relax.  In the Appendix, we show that this notion of ``distance'' along $G_\textrm{leq}$ can be expressed as the Kullback-Leibler divergence $D_\textrm{KL}$ between the local equilibrium density $\rho_{\beta,\beta_\textrm{b}}^\textrm{leq}$ given in Equation~\eqref{eq:leq} and the global equilibrium density $g_{\beta_\textrm{b}}$.  In particular, $D_\textrm{KL}[\rho_{\beta,\beta_\textrm{b}}^\textrm{leq}, g_{\beta_\textrm{b}}]$ is a monotonic function of  $a_\textrm{L}$ (defined in Equation~\ref{eq:leq1}), which is the natural parameter for the manifold $G_\textrm{leq}$.

Now we can understand how the (metastable) Mpemba effect can arise.  In the example shown in Figure~\ref{fig:alpha3plot}, the distance along $G_\textrm{leq}$ initially increases with $T$ and so does the total equilibration time.   But then this distance decreases for higher temperatures, leading to the Mpemba effect.  We note that in our approximation, the time to traverse the initial stage is much shorter than the time to relax along the green curve, so that variations in the length of the initial trajectory are irrelevant.  

If the bath temperature were changed at a finite rate (rather than a hot system being quenched directly into the bath), then the dynamics would be different.  For example, if the system is very slowly cooled from the initial temperature to final bath temperature, the trajectory would follow the quasi-static locus (red curve $G$) and no Mpemba effect would be possible.  Having shown that no Mpemba effect is possible with an infinitely slow quench and that the effect can be observed in the limit of an infinitely rapid quench, we can conclude that the Mpemba effect requires a \textit{sufficiently fast} temperature quench.

\vspace{1em}

\section{Metastable Mpemba effect in terms of extractable work}

\vspace{1em}

Our main goal is to express the criterion for the Mpemba effect in more physical terms.  For the metastable setting described above, we will find such a criterion in terms of a thermodynamic work.  We recall that the second law of thermodynamics for a system in contact with a single
thermal bath of temperature $\beta_\textrm{b}^{-1}$ can be expressed in terms of work and free energy rather than entropy:
\begin{align}
	W \geq \triangle F_{\textrm{neq},\beta_\textrm{b}} \,,
\label{eq:2nd-law}
\end{align}
where $W$ is the work received by the system and $\triangle F_{\textrm{neq}}$ denotes the difference in \textit{nonequilibrium free energies} (final $-$ initial values).  See, for example, \citet{gavrilov17b}, Equation~5 and associated references.

We recall also that the nonequilibrium free energy generalizes the familiar notion of free energy to systems out of equilibrium.  Thus, in analogy to Equation~\ref{eq:nrllibre}, we define
\begin{align}
	F_{\textrm{neq},\beta_\textrm{b}} \left( \rho \right) \equiv E (\rho) 
		-\beta_\textrm{b}^{-1} S\left( \rho \right) \,,
\label{eq:Fneq}
\end{align}
where the average energy $E(\rho)$ and Gibbs-Shannon entropy $S(\rho)$ are given by
\begin{align}
	E(\rho) &\equiv \int_{-\infty}^{+\infty} \dd{x} \rho(x) U(x)  \nonumber \\
	S(\rho) &\equiv -\int_{-\infty}^{+\infty} \dd{x} \rho(x) \ln \rho(x) \,.
\end{align}
These expressions reduce to their usual definitions for $\rho = g_{\beta_\textrm{b}}$ but can be evaluated, as well, over nonequilibrium densities.

In the formulation of the second law of Equation~\eqref{eq:2nd-law}, the initial and final states are arbitrary.  In our case, the initial state is the (approximate) local equilibrium reached at the end of Stage 1.  In the final state, the system is in equilibrium with the bath.

Physically $-\triangle F_{\textrm{neq}}$ represents the maximum amount of work that may be extracted from the nonequilibrium isothermal protocol \cite{parrondo15}.  We will refer to this quantity as the \textit{extractable work}.
\begin{align}
	W_\textrm{ex} \equiv - \triangle F_{\textrm{neq},\beta_\textrm{b}} \,.
\end{align}

\begin{figure}[h!]
\begin{center}
	\includegraphics[width=7cm]{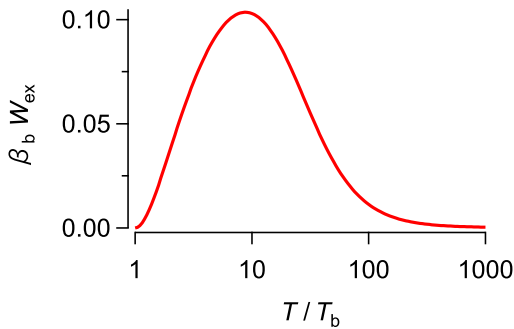}
\end{center}
\caption{Extractable work is a non-monotonic function of initial temperature $T=\beta^{-1}$ for the double-well potential of Figure~\ref{fig:localEq}A.}
\label{fig:extractableWork}
\end{figure}

In the Appendix, we show that the difference in nonequilibrium free energies $\triangle F_{\textrm{neq}}$ may be expressed as a Kullback-Leibler divergence.  Explicitly,
\begin{align}
	\triangle F_{\textrm{neq}} 
		&=  - \left[ F \left( \rho_{\beta,\beta_\textrm{b}}^\textrm{leq} \right) 
			- F \left( g_{\beta_\textrm{b}} \right) \right] \nonumber \\
		&=  - \beta_\textrm{b}^{-1} D_\textrm{KL} \left( \rho_{\beta, \beta_\textrm{b}}^\textrm{leq},
			g_{\beta_\textrm{b}} \right) \,.
\label{eq:beau}
\end{align}
In our set-up, the extractable work between the ``intermediate''
time (end of Stage 1) where $F_{\textrm{neq},\beta_\textrm{b}} = F_{\textrm{neq},\beta_\textrm{b}} \left( \rho_{\beta, \beta_\textrm{b}}^\textrm{leq} \right)$, and the final time of the slow evolution (where $F_{\textrm{neq},\beta_\textrm{b}} = F_{\textrm{eq}, \beta_\textrm{b}})$, is given by Equation~\eqref{eq:beau}: 
\begin{align}
	W_\textrm{ex} \left( \beta, \beta_\textrm{b} \right)
	=  \beta_\textrm{b}^{-1} D_\textrm{KL} 
		\left( \rho_{\beta, \beta_\textrm{b}}^\textrm{leq}, g_{\beta_\textrm{b}} \right)  \,.
\label{eq:WexDKL}
\end{align}
In Sec.~\ref{sec:metastable} and Figure~\ref{fig:alpha3plot}, we saw that $D_\textrm{KL}( \rho_{\beta, \beta_\textrm{b}}^\textrm{leq}, g_{\beta_\textrm{b}} )$ can be non-monotonic as a function of $\beta$.  We thus conclude that there can be a non-monotonic dependence on $\beta$ of the function
\begin{align}
	\beta \rightarrow W_\textrm{ex} \left( \beta,\beta_\textrm{b} \right) \,.
\label{eq:main-result}
\end{align}
This is our main result:  If the metastable Mpemba effect occurs, then the extractable work from the local-equilibrium state at the end of Stage 1 is non-monotonic in the initial temperature $\beta^{-1}$.  Figure~\ref{fig:extractableWork} shows an example, again calculated for the potential considered by \citet{kumar20}.

In addition to having a clear physical interpretation, $W_\textrm{ex} (\beta, \beta_\textrm{b})$ is easily calculated as a simple numerical integral of equilibrium Gibbs-Boltzmann distributions for two temperatures.  By contrast, to establish the non-monotonicity of $a_2$, the criterion of \citet{lu17}, one must first find the left eigenfunction $u_2$ by solving the boundary-value problem associated with the adjoint Fokker-Planck operator.

\section{Discussion}

\vspace{1em}

The anomalous relaxation process known as the Mpemba effect is defined by a non-monotonic dependence of relaxation time on initial temperature.  \citet{lu17} showed that an equivalent criterion is the non-monotonicity of the $a_2$ projection coefficient derived from an associated Fokker-Planck equation.  In this Brief Research Report, we have shown that, for a 1D potential $U(x)$ with a metastable and a stable minimum, the Mpemba effect can be viewed as a simple two-stage relaxation in the function space of all admissible probability densities.  In the fast Stage 1, the system relaxes to a local equilibrium.  In the slow Stage 2, the populations in the two coarse-grained states equilibrate.  In such a situation, we have shown that the Mpemba effect is associated with a  non-monotonic temperature dependence of the maximum extractable work of the local equilibrium stage reached at the end of Stage 1.Relative to the $a_2$ coefficient, extractable work is a much more physical quantity that is also much easier to calculate.

The physical picture offered here, for a double-well potential, meets our goal:  We can relate the existence of the Mpemba effect to a non-monotonicity of the extractable work.  However,  we have not carefully characterized the range of validity of the approximations used in our analysis.  For example, in writing Equation~\eqref{eq:leq}, we assume that the fraction of initial systems that start in either state ($x<0$ or $x>0$) is preserved after the initial fast transient.  In fact, even during the brief transient, calculating the fraction of systems in each region is subtle, a point emphasized by \citet{van-kampen77} in a careful study that would be the starting point for a more detailed theoretical investigation.

Although our arguments assume a 1D potential with two local states, they generalize easily to many dimensions and many local states.  In such cases, the state vector has a large number of dimensions, and solving the Fokker-Planck equation or even calculating its eigenfunctions is difficult. But calculating the extractable work remains easy.   Of course, our arguments do not imply that the Mpemba effect can occur \textit{only} in potentials with metastable states and leave open the possibility for other scenarios. 
%with anomalous, Mpemba-like relaxation.

% \red{Finally, the geometrical picture of the dynamics presented in Fig.~\ref{fig:alpha3plot} (and its three-dimensional counterpart in the supplement) invite one to consider the Mpemba effect from a more topological point of view. [DISCUSS BRAIDING?]. The main result arising from the view of braiding presented here or from the qualitatively similar notion of \textit{Mpemba index} presented in \citet{klich19} is a straightforward classification of types of Mpemba effects and an understanding as to why there is at least some robustness in the effect when conditions are slightly changed.  It would be interesting to see whether any other significant insights may be gained by pursuing such topological notions.}

%\newpage

\begin{widetext}

\section*{Appendix}
\label{sec:app}

\textit{1.  Monotonicity of Kullback-Leibler divergence along} $G_\textrm{leq}$.  The Kullback-Leibler divergence \cite{cover06} can be written in terms of Equation~\eqref{eq:leq} as
\begin{align}
	D_\textrm{KL} \left( \rho_{\beta,\beta_\textrm{b}}^\textrm{leq}, g_{\beta_\textrm{b}} \right)
	 &= \int_{-\infty}^\infty \dd{x}\rho_{\beta,\beta_\textrm{b}}^\textrm{leq} (x)
		\ln \left[ \frac{\rho^\textrm{leq}(x)}{g_{\beta_\textrm{b}}} \right] \nonumber \\
	&= \int_{-\infty}^0 \dd{x} a_\textrm{L}  \left( \frac{g_{\beta_\textrm{b}}(x)}
		{\int_{-\infty}^0 \dd{x'} g_{\beta_\textrm{b}}(x')} \right) 
		\ln \frac{a_\textrm{L} g_{\beta_\textrm{b}}(x)}
		{[{\int_{-\infty}^0 \dd{x'} g_{\beta_\textrm{b}}(x')}] \, g_{\beta_\textrm{b}}(x)} 
			+ \int_0^\infty \dd{x} \cdots \nonumber \\
	&= a_\textrm{L} \ln \left( \frac{a_\textrm{L}}{a_\textrm{L}^*} \right) + 
		a_\textrm{R} \ln \left( \frac{a_\textrm{R}}{a_\textrm{R}^*} \right) \,.  \nonumber \\
	&= D_\textrm{KL} \left[ \mqty(a_\textrm{L} \\ a_\textrm{R} ),
		\mqty(a_\textrm{L}^* \\ a_\textrm{R}^* ) \right] \,.
\label{eq:DKLlocal}
\end{align}
In the second line, we omit the corresponding $a_\textrm{R}$ terms.  In the third line, $a_\textrm{L}^* \equiv \int_{-\infty}^0 \dd{x} g_{\beta_\textrm{b}}(x)$ and $a_\textrm{R}^* \equiv \int_0^\infty \dd{x} g_{\beta_\textrm{b}}(x)$.  In the fourth line, the vectors represent two-state probability distributions.  Note that in the ``short Stage 1'' approximation of Equation~\eqref{eq:leq1}, the final expression for $D_\textrm{KL}$ involves two coarse-grained probability distributions, with $\smqty(a_\textrm{L} \\ a_\textrm{R} )$ depending only on $\beta$ and $\smqty(a_\textrm{L}^* \\ a_\textrm{R}^* )$ only on $\beta_\textrm{b}$.

We then investigate the monotonicity of $D_\textrm{KL} \left[ \mqty(a_\textrm{L} \\ a_\textrm{R} ),\mqty(a_\textrm{L}^* \\ a_\textrm{R}^* ) \right]$ by differentiating:
\begin{align}
	\dv{D_\textrm{KL}}{a_\textrm{L}} =  \ln \left( \frac{a_\textrm{L}}{a_\textrm{R}} \right) - 
	\ln \left( \frac{a_\textrm{L}^*}{a_\textrm{R}^*} \right) \,,
\label{eq:DKLlocal-deriv}
\end{align}
which is positive for $a_\textrm{L} > a_\textrm{L}^*$ and negative for $a_\textrm{L} < a_\textrm{L}^*$.  (Recall that $a_\textrm{L} + a_\textrm{R} = a_\textrm{L}^* + a_\textrm{R}^* = 1$.)  Thus, $D_\textrm{KL} \left( \rho^\textrm{leq}, g \right)$ is monotonic in $a_\textrm{L}$ on either side of equilibrium.

\noindent \textit{2.  Proof of Equation~\eqref{eq:beau}.} The relationship is well known \cite{shaw84} and holds for any distribution, including ones describing local equilibrium.  Below, to simplify notation, we write $\rho^\textrm{leq}$ for $\rho_{\beta,\beta_\textrm{b}}^\textrm{leq}$ and $g$ for $g_{\beta_\textrm{b}}$.
\begin{align*}
	D_\textrm{KL} \left( \rho^\textrm{leq}, g \right) 
	&= \int_{-\infty}^\infty \dd{x} \rho^\textrm{leq}(x)
		\ln \left[ \frac{\rho^\textrm{leq}(x)}{g(x)} \right] \\
	&= \int_{-\infty}^\infty \dd{x} \rho^\textrm{leq}(x)
		\ln \rho^\textrm{leq}(x) 
	- \int_{-\infty}^\infty  \dd{x} \rho^\textrm{leq}(x) 
		\ln g(x) \\
	&= -S \left( \rho^\textrm{leq}\right) 
		- \int_{-\infty}^\infty  \dd{x} \rho^\textrm{leq}(x) 
		\left[ -\beta_\textrm{b} U(x) + \beta_\textrm{b} 
		F \left( g \right) \right] \\
	&= -S \left( \rho^\textrm{leq}\right) + \beta_\textrm{b}
		\left[E \left( \rho^\textrm{leq} \right) \right] 
		- \beta_\textrm{b} F \left( g \right) \\
	&=  \beta_\textrm{b} \left[ F \left( \rho^\textrm{leq} \right)
		- F \left( g \right) \right] \,,
\end{align*}
which is equivalent to Equation~\eqref{eq:beau}.
\end{widetext}

\section*{Funding}
JB and AK were supported by NSERC Discovery and RTI Grants (Canada).  RC acknowledges support from the Pacific Institute for Mathematical Sciences (PIMS), the French Centre National de la Recherche Scientifique (CNRS) that made possible his visit to Vancouver and the project RETENU ANR-20-CE40-0005-01 of the French National Research Agency (ANR).

\end{document}